\documentclass{iaus}
\usepackage{graphicx}

\newcommand\hnot{$H_\circ$}
\newcommand\kms{km~s$^{-1}$}
\newcommand\msun{$M_\odot$}

\def\be{\begin{equation}}
\def\ee{\end{equation}}

\title[Scaling Relations of Disks] 
{On the Scaling Relations of Disk Galaxiess}

\author[Riccardo Giovanelli]   
{Riccardo Giovanelli}

\affiliation{Dept. of Astronomy, Cornell University, \\ Space Sciences Bldg.,
Ithaca, NY 14853, USA \\ email: {\tt rg39@cornell.edu}}

\pubyear{2013}
\volume{289}  
\pagerange{1-8}
\jname{Advancing the physics of cosmic distances}
\editors{R. de Grijs, Editor}
\begin{document}

\maketitle

\begin{abstract}
The physical background of scaling laws of disk galaxies is reviewed.
The match between analytically derived and observed scaling laws is
briefly discussed. Accurate modeling of the fraction of baryons that
end populating a disk, and the conversion efficiency of those into stars,
remains a challenging task for numerical simulations. The measurement 
of rotational velocity tends to be made with criteria of convenience
rather than through rigorous definition. And yet, the Tully-Fisher 
and the disk size vs. rotational velocity relations exhibit surprisingly
low scatter. Practical recipes (and costs) to optimize the quality of
template relations are considered.
\end{abstract}

\firstsection 
\section{Introduction}

Thirtyfive years have passed since Brent Tully and Rick Fisher advocated the use of the relation
between optical luminosity $L$  and 21cm HI linewidth $W$, previously discovered by Mort Roberts, as a
tool to obtain redshift-independent distances of spiral galaxies (Roberts 1969;
Tully \& Fisher 1977). Since then, the Tully-Fisher relation (hereafter TFR) has been
used extensively in the measurement of \hnot and that of deviations from smooth Hubble flow,
to distances approaching $z\simeq 0.1$. Over the range of linewidths $W>100$ \kms, typically used
in those applications, the TFR exhibits a tight power-law behavior, where the linewidth is to first 
order equal to twice the  maximum rotational velocity of the disk within galactocentric radii populated 
with  detectable HI, and $L$ a proxy for total mass within that radius. A TFR template
will have a slope of between 3 and 4 in $\log L$ vs. $\log W$ and a scatter between 0.14 and 0.18 dex
in $\log L$ (0.35 to 0.45 mag), which translates to a distance uncertainty for a single galaxy
between 15\% and 20\%. These values vary somewhat depending mostly on the adopted
optical band. The quality of a TFR template, i.e. the accuracy with which it can accurately 
provide a distance prediction, depends on a variety of parameters, including primarily those 
pertaining to the selection criteria for membership in the sample used to define that template 
and the reliability of the corrections applied to the observed quantities. Improvements
to that quality should result from a clear understanding of the sources of scatter and of the physical
basis of the TFR.

\section{Towards a Physical Understanding of the TFR}

{\bf Take 1}: Assume that disk mass and
light are exponentials of scale length $r_d$. The luminosity is then $L\propto r_d^2 I(0)$,
where $I(0)$ is the central disk surface brightness. The mass within radius $r$ is 
$M(<r)\propto r v_{rot}^2$, so if the rotation curve becomes flat due to a DM halo
truncated at $nr$ scale lengths, then $M_{tot}\propto n r_d v_{max}^2$ and 
\be 
L_d \propto a v_{max}^4 ~~~~  {\rm with} ~~~~ a=(M_d/L_d)^{-2} m_d^2 n^{-2}I(0)^{-2}
\ee 
where $m_d$ is the mass fraction of the disk. A rough match for the slope of the TFR 
is recovered, but much is contained in $a$.

{\bf Take 2}. \cite[Mo \etal (1998)], model the DM halo, within which the disk 
galaxy resides, as a singular isothermal sphere. Then the rotational velocity $v_c$ 
is the same at all radii, and the mass density at, and the mean density within, radius 
$r$ are
\be
\rho(r)={v_c^2 \over {4\pi G r^2}} {\rm ~~~~~ and ~~~~~~} {\bar\rho={3v_c^2\over{4\pi G r^2}}}
\label{eq:dens}
\ee
The critical density of the universe at redshift $z$ is
\be
\rho_{crit}= {3H^2(z)\over{8\pi G}}
\ee
so the radius within which the mean density is $200\rho_{crit}$  
is ~ $r_{200}=v_c/10H(z)$ and the mass within $r_{200}$ is 
\be
M_{halo}=v_c^2 r_{200}/G = v_{200}^3/10 G H(z);
\ee
Introducing the disk mass $M_{disk}$ and $m_d=M_{disk}/M_{halo}$ we get
\be
M_d = {m_d v_{200}^3\over 10~G~H(z)} \simeq 1.7\times10^{11}h^{-1} {\rm M}_\odot~
\Biggl({m_d\over 0.05}\Biggr)~\Biggl({v_{200}\over 250~{\rm km s}^{-1}}\Biggr)^3
\Biggl[{H(z)\over H_\circ}\Biggr]^{-1}  \label{Md}
\ee
This is starting to look familiar, but: 
\begin{itemize}
\item halos are not singular isothermal spheres;
\item disks' masses are not negligible and they can alter the total density profile; 
\item we measure disk luminosity, not disk mass; 
\item we don't measure $v_{200}$ but rather a circular(?) velocity at whatever 
radius Nature kindly leaves a tracer for us to measure.
\end{itemize}

{\bf Take 3}. We thus abandon the singular isothermal sphere model and, 
still following Mo \etal ~(1998), model the mass density more realistically 
with the NFW profile:
\be
\rho(r) = {\rho_{crit} ~\delta_\circ \over (r/r_s)(1+r/r_s)^2}
\ee
where $\delta_\circ=4\rho(r_s)/\rho_{crit}$ and $c=r_{200}/r_s$ is  the concentration index.
The mass within radius $r$ is then
\be
M(r) = 4\pi \rho_{crit}~\delta_\circ r_s^3 \Biggl[{1\over 1+ c(r/r_{200})} - 
1 + \ln (1+r/r_{200})\Biggr],  \label{Mnfwr}
\ee
the total halo mass is
\be
M_{halo}= 4\pi \rho_{crit}\delta_\circ r_s^3~\Bigl[\ln (1+c) - {c/(1+c)}\Bigr]
\ee
and the rotation curve is
\be
{v(r)\over v_{200}} = \Bigl[{1\over x} 
{{\ln(1+cx)-cx/(1+cx)}\over {\ln(1+c)-c/(1+c)}}\Bigr]^{1/2} \label{eq:vrotnfw}
\ee
where $x=r/r_{200}$ and $v_{200}=(GM_{halo}/r_{200})^{1/2}$.

\begin{figure}[t]		
\begin{center}
 \includegraphics[width=2.5in]{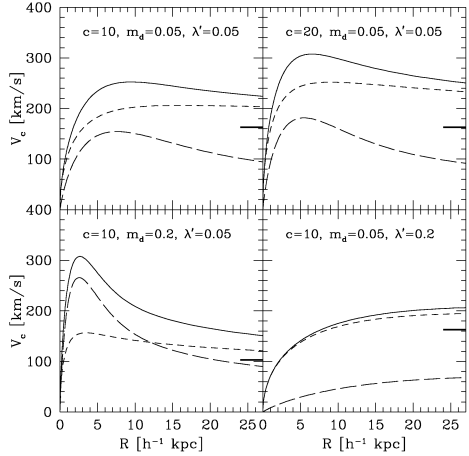} 
 \caption{
NFH rotation Curves of galaxies 
having the same disk mass ($5\times 10^{10} h^{-1}$ \msun)and different combinations of 
concentration indices, disk mass fractions, spin parameters, as inset within each box.
Disk and halo contributions to the rotation curve (solid line) are respectively shown as long-dash 
and short-dash lines. The thick line on the left of each panel is the circular velocity $v_{200}$
at the virial radius. Note the variance in the shape of the rotation curves and in their amplitudes
at any radius  (credit: Mo \etal ~1999).}
   \label{fig1}
\end{center}
\end{figure}

Figure 1 (after Mo \etal 1998) shows rotation curves of galaxies having the same disk mass 
but different combinations 
of concentration indices, disk mass fractions, spin parameters. The shape of 
the rotation curve, its maximum value and its value at the virial radius (indicated by the
short, thick line at the left of each box), vary widely from each other, illustrating the 
uncertainty that can be associated with the determination of the TFR linewidth parameter.
Because now $v_{rot}(r)$ is not constant, it matters at which radius $r_{meas}$ it is measured. 
It is sensible to measure it at a radius near which the gradient of $v_{rot}$ is small,
e.g. the radius within which 80\% of the light is produced, or a few disk scale lengths $nr_d$ 
out (a value of $n\simeq 3$ is often used, and $r_{80}$ is near $3r_d$). 
The equation for $M_d$ is then modified via a fudge factor which accounts for the shape of 
$V_{rot}$ and the radius at which it is measured. E.g. if $r_{meas}=3r_d$, then that fudge 
factor is $f_v=[v_{rot}(3r_d)/v_{200}]^{-3}$.
Finally, to convert the disk mass relation into a TFR look-alike, $M_d$ is divided by a stellar
mass--to--luminosity ratio; a star formation efficiency factor $\epsilon_{sf}$ is also introduced,
which accounts for the fraction of the disk baryon mass which has been converted to stars.
Thus:
\be
L_d =  1.7\times10^{11}h^{-1} {\rm M}_\odot ~ \epsilon_{sf}~
\Bigl({m_d\over 0.05}\Bigr)~\Bigl({M\over L}\Bigr)^{-1}_*~
\Bigl[{v_{obs}(3r_d)\over 250~{\rm km s}^{-1}}\Bigr]^3
\Bigl[{H(z)\over H_\circ}\Bigr]^{-1}f_V  \label{eq:tfr}
\ee
Accurately modeling $m_d \epsilon_{sf} (M/L)^{-1}_*$ is difficult, and so is measuring $f_v$. 
Moreover, the value of the rotation curve at $3 r_d$ may not be known from the kinematic data.
Much progress has been made since \cite[Steinmetz \& Navarro (1999)]{SteinNav99} in
modeling the gastronomy of disks, thanks to increased resolution of numerical simulations 
and a higher grade of sophistication in treating feedback
mechanisms, including adiabatic contraction of the halo (e.g. 
Governato \etal. 2007; Piontek \& Steinmetz 2011).
Generally, simulations recover the slope of the TFR but have problems in reproducing the
photometric zero-point. In particular, the stellar mass fraction in hydrodynamical simulations 
tends to be too high and bulges too large, while galaxies at low $z$ are too gas-poor  
(Scannapieco \etal 2012). Improvements are obtained by forcing a slow-down in the 
growth of bulges. However, processes of feedback, those affecting ISM structure and metallicity and their
effect on star formation take place on scales which are still much smaller than those 
currently achievable by cosmological simulations. As for $f_v$, an interesting observational 
result has recently come to the fore, throwing additional light on equation \ref{eq:tfr}, as 
we see next.

\cite[Reyes \etal. (2012)]{Rey12} directly measured averaged $f_v$, implementing a technique
earlier adopted by \cite[Seljak (2002)]{Sel02}. They used the SDSS data base,
stacking images of 133,598 disk galaxy images at a mean $z=0.07$, in three bins 
of stellar mass. The signature of weak lensing was identified in the stacked
images, thus yielding averaged halo masses out to $r_{200}$. The stellar mass
bins were centered respectively at $M_* = 0.6$, 2.7 and $6.5\times 10^{10}$ \msun. 
They could then measure $M_{200}$ and $r_{200}$, hence $v_{200}$. From a child catalog
of galaxies with resolved rotation curves, they obtained $v_{obs}(r_{80})$; then
the averages of the ratios $v_{obs}(r_{80})/v_{200}$, respectively for the three 
stellar mass bins, were found to be 1.27, 1.39 and 1.27. 
Interestingly, the corresponding stellar masses, as  fractions of the total 
baryon mass $f_{b} M_{200}$, computed assuming baryons are present with the 
cosmic abundance within the halo, are 0.15, 0.26 and 0.23 for the three stellar
mass bins, where $f_{b}=0.169$ is the cosmic baryon mass fraction. Disk galaxies
in the three stellar mass bins fall short of converting their baryons into stars by 
factors of between 4 and 5. This is illustrated graphically in figure 2 (after 
\cite[Papastergis \etal ~2012]{Pap12}). In each panel the abscissa is the halo mass;
in the left panel, the ordinate is the ratio of the stellar mass to $f_{b} M_{200}$.  
in the right panel, the ordinate is the ratio of the combined baryon mass of stars 
and cold disk gas to $f_{b} M_{200}$. 
The mismatch between the observed baryon mass in disks and that which would be
expected if the baryons were present within the halo in the fraction $f_{b}$
is thought to be mainly due: (a) for low halo masses: to the inability of the
halos to hold on to their baryons, which increases with decreasing halo mass;
(b) for galaxies near the peak of the curve: likely to the fact that in those 
systems most baryons are in hot galactic coronae, rather than in disks. The Reyes \etal (2012)
data are identified by the three circles near the top of the curve in the left panel.

Equation \ref{eq:tfr} and figure 2 summarize our understanding of the physical basis
of the TFR. It remains surprising that the variance on the combination of parameters
$m_d \epsilon_{sf} (M/L)^{-1}_* f_v$
is as small as indicated by the TFR observed scatter, and that the baryonic version of the TFR
remains a pure power law over 5 orders of magnitude in mass (see Figure 4 and discussion in Section 4).

\begin{figure}[t]		
\begin{center}
 \includegraphics[width=2.6in]{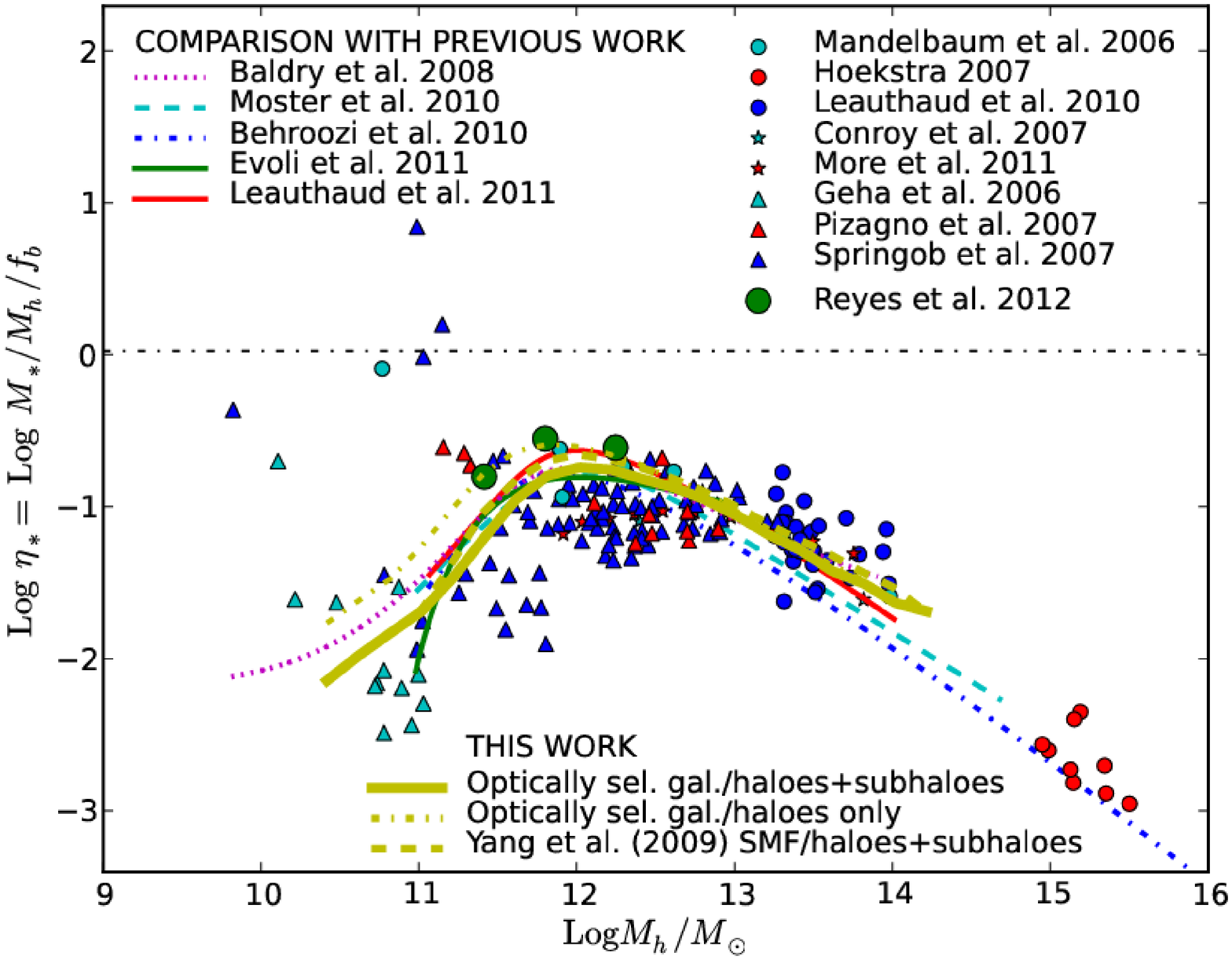} 
 \includegraphics[width=2.6in]{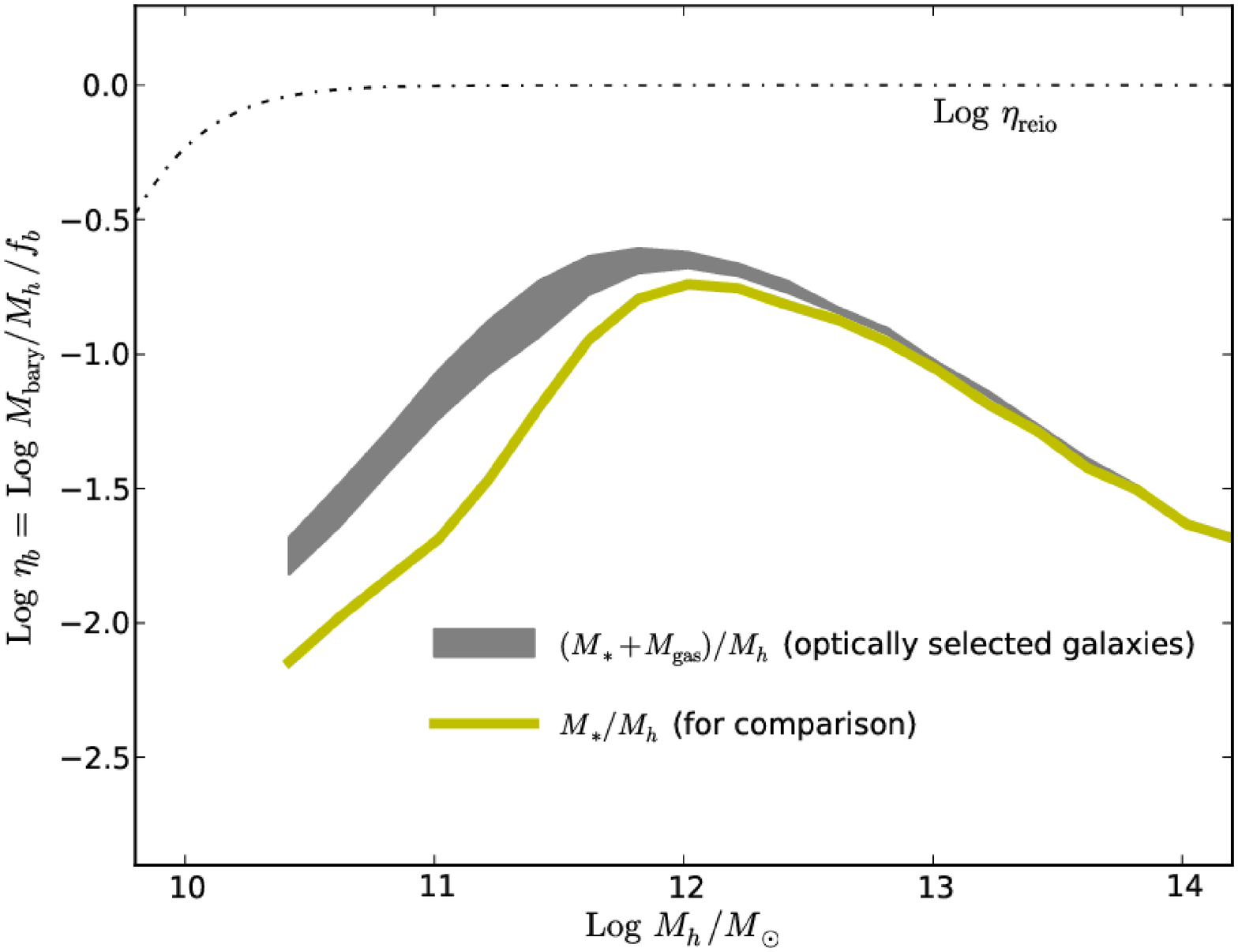} 
 \caption{{\it Left:} Stellar mass as a fraction of the baryon mass within the halo, 
plotted vs. the halo mass; the baryon mass within the halo is $f_b M_{halo}$, where $f_{b}$ is the cosmic baryon fraction. {\it Right:} The observable
baryon mass (stars plus disk ISM) as a fraction of the total baryon mass 
within the halo (shaded curve). The thin line is the stellar mass fraction, 
as in the panel to the left (credit: Papastergis \etal ~2012, reproduced with permission
off the American Astronomical Society.}
   \label{fig2}
\end{center}
\end{figure}

\section{Understanding Sources of Scatter: the RV Scaling Law}

An interesting case study on the
sources of scatter is that of the disk size vs. $V_{obs}$ (for short RV) scaling law.
Assuming the baryons collapse within the potential well of the halo  without altering its density 
profile, radiating but conserving angular momentum, and settling into a thin, exponential disk
of scale length $r_d$, a reasonment similar to that described in the previous section yields

\be
r_d \simeq
8.8h^{-1} \Bigl({\lambda\over 0.05}\Bigr)\Bigl({j_d\over m_d}\Bigr)
\Bigl({v_{c}\over 250~{\rm km~s}^{-1}}\Bigr)\Bigl[{H(z)\over H_\circ}\Bigr]^{-1}
 f_c f_R~~~{\rm kpc}
\label{RV}
\ee
where $\lambda$ is the spin parameter of the halo, $j_d$ the fraction of the angular momentum
carried by the disk, $f_c$ a fudge factor which depends solely on the concentration index of the
NFW halo profile and $f_R$ a factor which depends mainly  on the shape of the rotation 
curve. 

\begin{figure}[t]		
\begin{center}
 \includegraphics[width=3.2in]{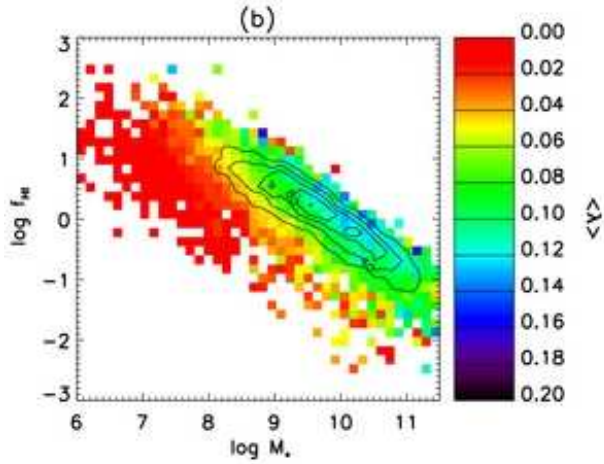} 
 \caption {HI gas fraction $f_{HI}=M_{HI}/M_*$ of 7459 ALFALFA survey galaxies
plotted aginst the stellar mass $M_*$; shading (color) represents the mean value of the
spin parameter $\lambda$ within each ($\log f_{HI},\log M_*)$ cell. Cells in the diagram 
containing more than 20 galaxies fall within the region bounded by contour lines. The typical 
galaxy used for TFR in cosmic distance studies falls within a few highly populated cells 
(those within the contour lines) in the diagram with $\log M_*>10$ and $\log f_{HI}>0.1$.
As Huang \etal ~point out, the mean value of $\lambda$ remains nearly constant along lines
of constant $\log f_{HI}$. See section 3.1 for discussion. (credit: Huang \etal ~2012.
Reproduced with permission from the American Astronomical Society)}
\label{fig3}
\end{center}
\end{figure}
An important source of scatter in this relation would appear to be that driven by the 
variance in $\lambda$. N-body simulations show that, over all halos, $\lambda$ has a 
log-normal distribution centered near 0.045
with a scatter of 0.22 dex in $\log \lambda$. Thus, it would be expected that the observed
scatter of the RV law would be greater than 0.22 dex in $\log r_d$.  Yet, the observed
scatter is no greater than 0.16 (\cite[Courteau \etal ~2007]{Cour07}, \cite[Saintonge \&
Spekkens ~2011]{SS11}). The possible explanation of the mismatch is in Figure 14 of 
\cite[Huang \etal ~(2012)]{huang12}, reproduced here as figure 3. Using the $\alpha.40$
release pf the ALFALFA survey (\cite[Haynes \etal ~2012]{alpha.40}), Huang \etal ~derive spin
parameters of 7459 galaxies using the estimator 
\be
\lambda = 21.8 {r_d [{\rm kpc}] \over V_{rot}^{3/2}[{\rm km ~s}^{-1}]}. \label{eq:lambda}
\ee
The inferred values of $\lambda$ are very coarse, and their dispersion is, as a result, 
much broader than the intrinsic one. In a plane of HI gas fraction $f_{HI}=M_{HI}/M_*$ 
vs. stellar mass $M_*$, the average value of $\lambda$ within each cell of that plane
is represented by a different degree of shading (color). Cells containing more than 20 
galaxies fall within the region bounded by contour lines. The typical galaxy used by the 
TFR in cosmic distance studies falls within a few highly populated cells (those within 
the contour lines) of the diagram, stretching along lines of roughly constant HI mass  
in the region with $\log (M_*/M_\odot)>9.5$ and higher than average $\log f_{HI}$ for any given $M_*$. 
Because the $\lambda$ estimator 
used is a ragged one, it is not possible to derive the intrinsic $\log \lambda$ scatter.
However, as Huang \etal ~point out, it is apparent in figure 3 that the mean value of 
$\lambda$ remains nearly constant along lines of constant $M_{HI}$. Thus, the typical 
TFR target galaxies are extracted from a sample 
with much tighter variance than the HI-selected lot shown in the figure. It appears safe 
to assume that the scatter in $\log \lambda$ in galaxies of a TFR sample is smaller by
at least a factor of 2 than that in any sample representative of the global halo mass function 
resulting from numerical simulations, estimated to be about 0.22 dex. 
The less than initially feared impact of $\lambda$-scatter on the RV law makes the latter a 
desirable complement to TFR. \cite[Saintonge \& Spekkens (2011)]{SS11} find that the scatter 
of the I--band RV relation can be reduced to $\sim 0.11$ dex on the linear size parameter,
if $r_d$ is replaced with an isophotal radius measured
at a large enough galactocentric distance so that the disk is transparent and the isophotal
radius requires no correction for opacity. Such an RV relation is comparable in predictive 
quality with the TFR. Saintonge \& Spekkens use a template RV relation using the Cornell SCI 
data set to derive an estimate of the Hubble parameter: \hnot $=72\pm 7$ \kms~Mpc$^{-1}$.

\section{Ways to Improve Your Scaling Law Template}

Significant benefits to a cosmic distances program based on disk galaxy scaling laws 
can result from wise choices in terms of template sample selection and optimization
of photometric and kinematic parameters.
\begin{figure}[b]		
\begin{center}
 \includegraphics[width=4.0in]{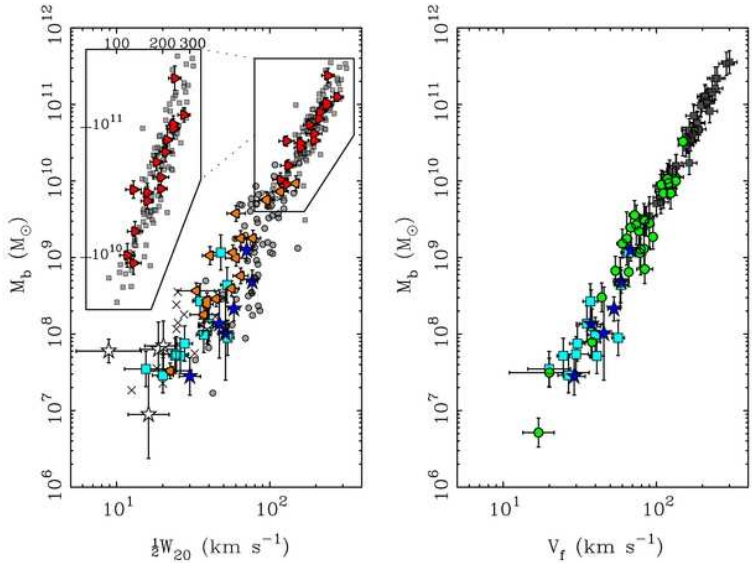} 
 \caption {Baryonic mass as a function of half line width (left) and circular velocity
as obtained from resolved rotation curves (right). The section of the plot  for the
high mass end of the relation is overplotted with higher resolution on the left panel.
Note (a) the power law extension of the relation all the way to the low mass end 
and (b) the significant reduction of the scatter when the rotational velocity
is extracted from a resolved rotation curve (credit: McGaugh ~2012; reproduced 
with permission from the author and the American Astronomical Society).}
\label{fig4}
\end{center}
\end{figure}

\begin{itemize}
\item {\bf Sample Selection: All Sky.}
The CMB dipole is due to the peculiar velocity of the Milky Way. It
produces an apparent reflex motion, with respect to the observer, of galaxies in 
shells of progressively increasing radius. Asymptotically, that motion 
converges both in amplitude and apex direction toward the CMB dipole vector.
Convergence is reached at shell radii of $\sim 100$ Mpc (\cite
[Giovanelli \etal ~1998]{RG98}, \cite[Dale \& Giovanelli 2000]{DG2000}). 
Combined with their clustering properties, an uneven distribution of 
galaxies in the template sample can produce TFR template relations with an incorrect
zero-point, simulating inflating or deflating, Local Group-centered Hubble bubbles.
\item {\bf Sample Selection: Basket of Clusters. }
Photometric parameters of galaxies, such as inclinations to the line of
sight and disk scale lengths, are best measured for relatively nearby objects. 
Peculiar velocities of galaxies can contribute a measurable fraction of
TFR scatter at distances up to $\simeq 100$ Mpc. In the case of a cluster with
N measured members, the offset from a template due to the cluster motion
can be gauged with an accuracy about $\sqrt N$ times better than for a
single galaxy. Thus, if the TFR template sample consists of galaxies in 
clusters, the component of the scatter introduced by each cluster's peculiar 
velocity is removable. This can be achieved by adding cluster galaxies to 
the template after their offset from the mean TFR, due to the peculiar 
velocity of their parent cluster, has been removed. A template relation built 
from galaxies in a basket of clusters will thus have smaller scatter than
one obtained for a field galaxy sample. 
\item {\bf Photometric Band.}
Because a good estimate of the kinematical parameter in the TFR requires
that galaxies be sufficiently inclined to the line of sight, corrections
to the optical flux for internal extinction are an important source of
TFR scatter. This advises for selecting a photometric band within which
internal extinction corrections are small, a concern that needs to be weighed 
against the requirement that the photometry be as representative as possible 
of the stellar population. Major TFR surveys of the last two  decades
chose either R or I optical bands as a compromise between the two requirements. 
More recently, the wide field coverage of the {\it Spitzer Space Telescope} archival 
data in the 3.8 $\mu$m band has offered the opportunity to build substantial TFR 
samples for which extinction corrections --- very small --- and homogeneity of the 
photometric source offer the possibility of reduced scatter. The recent work of
Sorce \etal ~(2012) indicates that, while the reliance in a consistent 
photometric scale over the whole sky is a definite plus, improvements in TFR 
scatter by going to the mid--IR  are not yet substantial: the scatter of a 
preliminary 3.8 $\mu$m template is 0.49 mag, which can be reduced to 0.42 mag 
after introducing a color correction. Contamination by emission from hot dust 
and PAHs may play a role in the relatively high scatter of this relation.

As just indicated, a color correction can have an important impact in reducing 
the TFR scatter. Until recently, large TFR samples were monochromatic. A dependence
of the TFR template attributed to morphological type has often been found in
monochromatic TFR templates (e.g. figures 3 and 4 of Giovanelli \etal ~1997): that 
is in fact a color dependence. Availability of multicolor photometry can thus be a 
good asset towards building a good scaling relation template.
\item {\bf Stellar vs. Baryonic.}
In the TFR, disk luminosity is considered a proxy for stellar mass, and the relation 
becomes significantly noisier for disks with rotational velocities smaller than 50 \kms.
However, the power law character of the TFR is preserved through slower
rotators if disk luminosity is replaced with the baryon mass, defined as the sum of
the ISM mass, as obtained via the 21cm HI line, and the stellar mass as
inferred from the disk luminosity and color. Figure 4 (after McGaugh 2012)
shows the baryonic TFR  behavior as a single power law to
rotational velocities as low as 20 \kms. The usefulness of this
relation can apply to the determination of distances to nearby dwarf systems, as an
economical substitute for techniques such as the measurement of the TRGB.
\item {\bf Line Widths vs. Resolved Rotation Curves.}
The two panels of figure 4 differ in the adoption of the kinematic parameter: the scatter
in the baryonic TFR is significantly reduced if the rotational velocity is inferred from
a resolved rotation curve (the label $V_f$ indicates that the velocity was measured at 
a radius at which the rotation curve becomes "flat"). A template that uses single-dish HI 
line widths is economical,
but it can be improved if replaced with H$\alpha$ rotation curves and, even more so, if HI
synthesis maps  are available, for the HI can be  typically detected to larger galactocentric
radii than H$\alpha$ and therefore provide a better sampling of the large-scale dynamics of the 
galaxy.
\item{\bf The Cost of Reduced Scatter.}
Several of the recommendations in this section require more data than does the
''basic" TFR. For the added value, there is a cost penalty. Since we live in times of fiscal 
austerity, it can be useful to quantify the cost of reduced scatter. The more "expensive" 
observational parameter is the kinematical one. Consider three versions of the latter, 
more often used, and the operating costs of telescopes needed to obtain them., and see how 
they would impact a TFR project budget:
\begin{itemize}
\item Single dish HI line widths; 
cost: $\sim \$ 200$ gal$^{-1}$; feasible sample size: $\sim 10^4$
\item H$\alpha$ rotation curve;
 cost: $\sim \$ 2$K gal$^{-1}$; feasible sample size: $\sim 10^3$
\item HI synthesis map; 
cost: $\sim \$ 15$K gal$^{-1}$; feasible sample size: $\sim 10^2$
\end{itemize}
These are very rough numbers. Perhaps better deals can be had after careful shopping.
\item {\bf Invitation to Harvest .}
By the way, the ALFALFA HI survey is now complete and nearly half of the extragalactic HI sources 
are in the public domain. The final catalog will contain more than 30,000 sources, 
including optical identifications, HI masses and line widths. See Giovanelli \etal ~(2005)
and Haynes \etal ~(2011).
\end{itemize}

This work has been supported by a grant of the U.S. National Science Foundation.

\end{document}